\documentclass[a4paper, abstract, 11pt, twoside, headinclude]{scrartcl}

\usepackage[T1]{fontenc}
\usepackage[utf8]{inputenc}
\usepackage{lmodern}
\usepackage[english]{babel}
\usepackage{authblk}
\usepackage[headsepline]{scrlayer-scrpage}
\usepackage{standalone}
\usepackage{hyperref}
\usepackage{mathtools}
\usepackage{geometry}
\usepackage{microtype}
\usepackage{nicefrac}
\usepackage{enumitem}
\usepackage{tikz, pgfplots}
\pgfplotsset{compat=1.14}
\usetikzlibrary{positioning, patterns}

\PassOptionsToPackage{noabbrev, capitalise}{cleveref}

\input{abbrv}
\input{thmdef}

\usepackage{xcolor, xspace}

\setkomafont{date}{\normalsize}

\setlist[description]{font={\normalfont\scshape}}
\setlist[itemize]{label={\raisebox{1pt}{\tiny$\blacksquare$}}}

\title{The Computational Complexity of Understanding Network Decisions}
\author[1]{Stephan W\"{a}ldchen}
\author[1]{Jan Macdonald}
\author[1]{Sascha Hauch}
\author[1,2]{Gitta Kutyniok}
\affil[1]{Institut f{\"u}r Mathematik, Technische Universit{\"a}t Berlin}
\affil[2]{Fakult{\"a}t Elektrotechnik und Informatik, Technische Universit{\"a}t Berlin}
\affil[ ]{\texttt{\{stephanw, macdonald, hauch, kutyniok\}@math.tu-berlin.de}}

\makeatletter
\rohead{S. W\"{a}ldchen, J. Macdonald, S. Hauch, G. Kutyniok}
\lehead{\@title}
\makeatother

\begin{document}
\maketitle

\begin{abstract}
    For a Boolean function $\Phi\colon\skl{0,1}^d\to\skl{0,1}$ and an assignment to its variables  $\bfx=(x_1, x_2, \dots, x_d)$ we consider the problem of finding the subsets of the variables that are sufficient to determine the function value with a given probability $\delta$. This is motivated by the task of interpreting predictions of binary classifiers described as Boolean circuits (which can be seen as special cases of neural networks). \par 
    We show that the problem of deciding whether such subsets of relevant variables of limited size $k\leq d$ exist is complete for the complexity class $\SNP^{\SPP}$ and thus generally unfeasible to solve.
    We introduce a variant where it suffices to check whether a subset determines the function value with probability at least $\delta$ or at most $\delta-\gamma$ for $0<\gamma<\delta$. This reduces the complexity to the class $\SNP^{\SBPP}$. \par 
    Finally, we show that finding the minimal set of relevant variables can not be reasonably approximated, i.e. with an approximation factor $d^{1-\alpha}$ for $\alpha > 0$, by a polynomial time algorithm unless $\SP = \SNP$ (this holds even with the probability gap).
\end{abstract}

\section{Introduction}\label{sec:introduction}

Algorithmic problem solving in real-world scenarios often requires reasoning in an uncertain environment. This necessity lead to the investigation of probabilistic satisfiability problems and probabilistic computational complexity classes such as $\SPP$ and $\SNP^{\SPP}$.
One prototypical example, the \textsc{E-Maj-Sat} problem \cite{littman1998computational, littman2001stochastic}, is an extension of the classical satisfiability problem that includes an element of a model counting formulation. 
The class of $\SNP^{\SPP}$-complete problems contains many relevant artificial intelligence (AI) problems such as probabilistic conformant planning \cite{drummond1990anytime, kushmerick1995algorithm}, calculating maximum expected utility (MEU) solutions \cite{dechter1998bucket}, and maximum a posteriori (MAP) hypotheses \cite{park2002map}.

\begin{figure}
    \centering
\begin{tikzpicture}[
  level distance=1.2cm,
  level 1/.style={sibling distance=1cm},
  level 2/.style={sibling distance=2cm},
  level 3/.style={sibling distance=1.2cm},
  edge from parent/.style={draw, shorten <=.1cm, <-, >=stealth},
  var/.style={},
  logic/.style={draw, rectangle},
] 

\node[var] (out) {$\Phi(x_1,x_2,x_3)$}
  child { node[logic] {$\odr$}
    child { node[logic] {$\und$} 
      child { node[var] {$x_1$} }
      child { node[var] {$x_2$} }
    }
    child { node[logic] {$\lnot$} 
      child { node[var] {$x_3$} }
    }
  };
\end{tikzpicture}
\hspace{1.5em}
\begin{tikzpicture}[
  var/.style={minimum width=.7cm},
  label/.style={minimum width=.7cm, font=\scriptsize},
  hidden/.style={draw, circle, minimum width=.7cm, font=\scriptsize, inner sep=0cm},
  edge/.style={shorten >=.1cm, ->, >=stealth},
]
  \node[hidden] (out) {$+1$};
  \node[below=.5cm of out, hidden] (or) {$+1$};
  \node[below left=.5cm and .2cm of or, hidden] (and) {$-1$};
  \node[below right=.5cm and .2cm of or, hidden] (not) {$+1$};
  \node[below left=.5cm and .1cm of and] (x1) {$x_1$};
  \node[below right=.5cm and .1cm of and] (x2) {$x_2$};
  \node[below right=.5cm and .1cm of not] (x3) {$x_3$};
  
  \draw[edge] (x1) -- (and) node[midway, left, label] {$+1$};
  \draw[edge] (x2) -- (and) node[midway, right, label] {$+1$};
  \draw[edge] (x3) -- (not) node[midway, right, label] {$-1$};
  
  \draw[edge] (and) -- (or) node[midway, left, label] {$-1$};
  \draw[edge] (not) -- (or) node[midway, right, label] {$-1$};
  
  \draw[edge] (or) -- (out) node[midway, right, label] {$-1$};
  
  \setlength\arraycolsep{2pt}
  \node[right=1cm of or, font=\scriptsize] (formula) {$-\varrho\kl{\begin{bmatrix} -1 & -1 \end{bmatrix}\varrho\kl{\begin{bmatrix} 1 & 1 & 0 \\ 0 & 0 & -1 \end{bmatrix} \begin{bmatrix} x_1 \\ x_2 \\ x_3 \end{bmatrix} + \begin{bmatrix} -1 \\ 1 \end{bmatrix}}+1}+1$};
\end{tikzpicture}
    \caption{The Boolean function $\Phi(x_1,x_2,x_3) = \kl{x_1\und x_2}\odr\kl{\lnot x_3}$ viewed as a Boolean circuit (left) and a rectified linear unit (ReLU) neural network in its graphical (middle) and algebraic representation (right). Network weights and biases are denoted at the edges and nodes respectively. The ReLU activation $\varrho(x) = \max\skl{x, 0}$ is applied components-wise.}
    \label{fig:circuit_nn}
\end{figure}
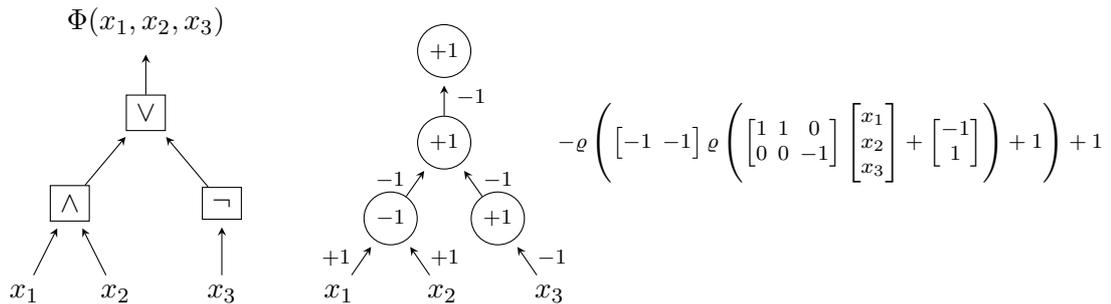

We connect these probabilistic reasoning tasks to an important problem in machine learning, namely the problem of interpreting the decisions of neural network classifiers.
Neural networks are parameter-rich and highly non-linear models and can be seen as continuous generalisations of Boolean circuits. This is briefly visualised in \cref{fig:circuit_nn}.
They have achieved impressive success in classification \cite{ieee6638947speech, NIPS2012_4824, NIPS2013_5207} and regression tasks \cite{sun2013deep, toshev2014deeppose, taigman2014deepface} and are increasingly also used to solve various inverse problems \cite{Kang2017lowdosect, Xie2012inpainting}.

However, the same expressiveness that allows for hierarchical reasoning and universal approximation makes understanding and interpreting these models more challenging compared to traditional machine learning methods like linear regression or decision trees.
Treating neural networks as ``black box'' solvers without accessible reasoning is not feasible in many circumstances, for example in critical applications such as medical imaging and diagnosis \cite{ MCBEE20181472, annurev-bioeng-071516-044442}.

A significant first step towards understanding network decisions is to distinguish the relevant input parameters from the less relevant ones for a specific prediction. This goal has been pursued predominantly for image classification problems in the form of visual maps that assign importance values to the inputs variables, for example in \cite{bach-plos15, erhan2009visualizing, simonyan2013deep, zeiler2014visualizing}.

We formalise this notion as a probabilistic decision problem in the following sense. Given a Boolean function and an assignment to its variables, is there a small set of input variables that if held constant determines the function value for almost all possible assignments to the rest of the variables?

This problem should be contrasted with the feature selection problem, where the goal is to find a subset of  globally important variables, independent from a concrete input. Instead, here we are interested in finding the important variables only for one specific input assignment.

We formulate the decision and minimisation version of the problem at hand and compare it to related problems in \cref{sec:problem}. We then analyse its computational complexity in \cref{sec:complexity} and prove that it is $\SNP^{\SPP}$-complete.
We propose a more practically relevant variant of the problem with a gap promise that allows for efficient computations of expectation values in \cref{sec:probvariants}. Analysing its complexity will finally show that any reasonable approximation algorithm to the minimisation version of the problem cannot be polynomial time unless $\SP=\SNP$.

\paragraph{Notation}
Throughout the paper $d\in\N$ denotes the arity of the considered Boolean function $\Phi\colon\skl{0,1}^d\rightarrow\skl{0,1}$ and $\bfx = \kl{x_1,\dots,x_d}\in\skl{0,1}^d$ is an arbitrary fixed assignment to its variables for which we are interested in finding the subsets of important variables. We denote the $d$-dimensional vectors of all zeros or ones by $\bfzero_d$ and $\bfone_d$ respectively. The function $\Phi$ is assumed to be described in terms of standard logical operations like AND, OR, and NOT. The description length of $\Phi$ is thus equal to the description length of the logical expression describing it. We denote $\ekl{d}=\skl{1,\dots,d}$ and for a subset $S\subseteq[d]$ denote by $\bfx_S = (x_i)_{i\in S}$ the restriction of $\bfx$ to components indexed by $S$. Further, we will use Boolean functions also interchangeably as logical propositions, in the sense that $\Phi(\bfx)$ is shorthand for the logical proposition $\Phi(\bfx)=1$. Whenever we talk about statements concerning probabilities of logical propositions to hold we assume independent uniform distributions for all involved variables. Thus, we have for example 
\[
 P_{\bfy}(\Phi(\bfy)) = \frac{\bkl{\skl{\,\bfy \in \skl{0,1}^d\,:\,\Phi(\bfy)=1\,}}}{\bkl{\skl{\,\bfy \in \skl{0,1}^d\,}}}.
\]
We omit the subscript whenever it is clear from the context over which variables the probability is taken. If the probability is taken over all variables of a Boolean function we simply write $P(\Phi)$ instead of $P_\bfy(\Phi(\bfy))$.

\section{Problem Formulation}\label{sec:problem}
Intuitively, a subset $S\subseteq \ekl{d}$ of variables is \emph{relevant} for the function value $\Phi(\bfx)$ if fixing $\bfx$ on $S$ and randomising it on the complement $S^c$ does not change the value of $\Phi$ with high probability. The complement then consists of the \emph{non-relevant} variables.
\begin{definition}
Let $\Phi\colon \skl{0,1}^d\to \skl{0,1}$, $\bfx \in \skl{0,1}^d$, and $\delta \in \ekl{0,1}$. We call $S\subseteq\ekl{d}$ a \emph{$\delta$-relevant} set for $\Phi$ and $\bfx$, if
\[
 P_{\bfy}\kl{\Phi(\bfy) = \Phi(\bfx)\,\middle|\,\bfy_S = \bfx_S } \geq \delta.
\]
\end{definition}
For $\delta$ close to one this means that the input $\bfx$ supported on $S$ already determines the output $\Phi(\bfx)$ with high probability. It is clear that $S = [d]$ is always $1$-relevant and any subset $S\subseteq\ekl{d}$ is $0$-relevant. Now the interesting question arises if for a given $\delta$ there exists a $\delta$-relevant set of a certain maximal size. Similarly, one could ask to find the smallest $\delta$-relevant set. This set would then be composed of the most important variables for the function value $\Phi(\bfx)$. There is an obvious ``rate-distortion'' trade-off in the sense that generally a larger $\delta$ will require a larger set $S$.

\begin{definition}
  For $\delta\in (0,1]$ we define the \textsc{Relevant-Input} problem as follows.
  \begin{description}
    \item[Given:] $\Phi\colon\skl{0,1}^d\to\skl{0,1}$, $\bfx\in\skl{0,1}^d$, and $k\in \N$, $1\leq k\leq d$.
    \item[Decide:] Does there exist  $S \subseteq\ekl{d}$ with $\bkl{S}\leq k$ such that $S$ is $\delta$-relevant for $\Phi$ and $\bfx$?
  \end{description}
\end{definition}

The minimisation formulation of the above decision problem can be defined in the obvious way.

\begin{definition}
   For $\delta\in (0,1]$ we define the \textsc{Min-Relevant-Input} problem as follows.
  \begin{description}
    \item[Given:] $\Phi\colon\skl{0,1}^d\to\skl{0,1}$ and $\bfx\in\skl{0,1}^d$.
    \item[Minimize:] $k\in\N$ such that there exists $S\subseteq\ekl{d}$ with $\bkl{S}\leq k$ and $S$ is $\delta$-relevant for $\Phi$ and $\bfx$.
  \end{description}
\end{definition}

The majority of the remainder of the paper will deal with analysing the computational complexity of the \textsc{Relevant-Input} and \textsc{Min-Relevant-Input} problems and related variants thereof. Before, we want to point out some similarities to a problem from cooperative game theory.

\subsection{Related Works}
Another concept for measuring the relevance or the \emph{contribution} of individual variables to a collective are the Shapley values \cite{shapley1952gamevalues} in cooperative game theory. Here the variables are seen as players of a coalitional game and the Shapley values describe a method to distribute the value achieved by a coalition of players to the individual players. This distribution fulfils a set of game theoretic properties that make it ``fair''.

Let $\nu\colon 2^{[d]} \to \mathbb{R}$ be a function that assigns a value to each subset of variables (coalition of players). It is called the \emph{characteristic function} of the game. Then the Shapley value of the $i$-th variable ($i$-th player) is defined as
\[
    \varphi_{i,\nu} = \sum_{S \subseteq [d]\setminus\{i\}} \frac{\bkl{S}!(d - \bkl{S} - 1)!}{d!} \kl{ \nu(S \cup \{i\}) - \nu(S) },
\]
which can be interpreted as the marginal contribution of the $i$-th variable to the value $\nu$ averaged over all possible coalitions. In general it is $\SSP$-hard to compute Shapley values \cite{deng1994complexity}. However, in some cases efficient approximation algorithms exist \cite{fatima2008shapley-approx}.

In our scenario the value of a subset of variables $S$ can be measured by the expected difference in $\Phi$ when fixing variables in $S$ and randomising the remaining variables. In \cite{kononenko2010efficient} it was proposed to use
\[
   \nu(S) = \frac{1}{2^{d-|S|}} \sum_{ 
   \substack{
   \bfy \in \skl{0,1}^d \\ \bfy_S = \bfx_S
   }
   } \Phi(\bfy)-\E_\bfy\kl{\Phi(\bfy)} 
\]
for the analysis of classifier decisions, which uses the expectation of the completely randomised classifier score as a reference value to determine the coalition value. We observe that
\begin{align*}
 P_{\bfy}\kl{\Phi(\bfy) = \Phi(\bfx)\,\middle|\,\bfy_S = \bfx_S } &= 1-\frac{1}{2^{d-|S|}} \sum_{ 
   \substack{
   \bfy \in \skl{0,1}^d \\ \bfy_S = \bfx_S
   }
   } | \Phi(\bfy)-\Phi(\bfx)| \\ 
   &= 1- |\nu(S)+\E_\bfy\kl{\Phi(\bfy)}-\Phi(\bfx)|,
\end{align*}
hence $S\subseteq\ekl{d}$ is $\delta$-relevant for $\Phi$ and $\bfx$ exactly if $|\nu(S)+\E_\bfy\kl{\Phi(\bfy)}-\Phi(\bfx)|\leq 1-\delta$.

Despite this relation between $\delta$-relevant sets and the characteristic function $\nu$ our problem formulation is considerably different from the Shapley value approach. The task considered in this paper is not to distribute the value of coalitions amongst the variables but to find (small) coalitions that are guaranteed to have a certain value. We will see that unlike for for Shapley values no efficient approximation algorithms can exist for this problem unless $\SP=\SNP$.

\section{Computational Complexity Analysis}\label{sec:complexity}

The first main theorem shows that the \textsc{Relevant-Input} problem  is generally hard to solve for $\delta\in\left[\frac{1}{2},1\right)$.
\begin{theorem}\label{thm:npppcomplete}
  For $\delta\in\left[\frac{1}{2},1\right)$ the \textsc{Relevant-Input} problem is $\SNP^{\SPP}$-complete.
\end{theorem}

 The proof of \cref{thm:npppcomplete} will be split into two parts. We will show that \textsc{Relevant-Input} is $\SNP^\SPP$-hard in \cref{sec:nppphard} and that it is contained in $\SNP^\SPP$ in \cref{sec:npppcontained}. Before we continue to give the proof we want to make a few more observations and remarks.

Intuitively, the $\SNP$-part of the problem complexity comes from the necessity to check all subsets $S\subseteq\ekl{d}$ as possible candidates for being $\delta$-relevant. The $\SPP$-part of the complexity comes from the fact that for any given set $S$ checking if it is $\delta$-relevant is by itself a hard problem (in fact $\SPP$-hard)\footnote{Checking if a subset is $1$-relevant is in $\ScoNP$ instead of $\SPP$. Thus we excluded $\delta=1$ in \cref{thm:npppcomplete}.}. The problem class $\SNP^\SPP$ is beyond the scope of conventional computing. In particular, \textsc{Min-Relevant-Input} is at least as hard to solve as the corresponding decision problem, which makes it unfeasible to solve exactly.  However, in applications it is rarely required to exactly find the smallest relevant set. It would be desirable to obtain good approximate solutions within feasible computational complexity. \par 
There are two potential ways for simplifying the problem by allowing approximations: Firstly, approximating the size of the minimal relevant set itself, and secondly relaxing the requirement that the set has to be exactly $\delta$-relevant. The former would address the $\SNP$ part whereas the latter would address the $\SPP$ aspect. \par 
Calculating probabilities or expectation values may be hard in theory, yet it is often easy to calculate them (approximately) in practice, for example by sampling. Checking whether a logical proposition is satisfied with probability more than $\delta$ this way only fails if the true probability can be arbitrarily close to $\delta$. These edge cases cause the hardness of the problem but in our scenario we do not necessarily care about their resolution. We discuss in \cref{sec:gapped} that this can be relaxed formally by stating a promise problem. This reduces the problem complexity from $\SPP$ to $\SBPP$. Unfortunately, even in this simplified case, it remains $\SNP$-hard to approximate the size of the optimal set $S$ to within any reasonable approximation factor. This is discussed in \cref{sec:inapproximability}.

\subsection{\textsc{Relevant-Input} is \texorpdfstring{$\SNP^\SPP$-hard}{NP\^{}PP-hard}}\label{sec:nppphard}
We now give the first part of the proof of \cref{thm:npppcomplete}. This is done by constructing a polynomial-time reduction of a $\SNP^\SPP$-complete problem to \textsc{Relevant-Input}. The canonical complete problem for $\SNP^\SPP$ is \textsc{E-Maj-Sat} \cite{littman1998computational}.
\begin{definition}
  The \textsc{E-Maj-Sat} problem is defined as follows.
  \begin{description}
    \item[Given:] $\Phi\colon\skl{0,1}^d\to\skl{0,1}$ and $k\in\N$, $1\leq k\leq d$.
    \item[Decide:] Does there exist $\bfx\in\skl{0,1}^k$ such that $P_{\bfy}\kl{\Phi(\bfy)\,\middle|\, \bfy_{[k]} = \bfx} > \frac{1}{2}$?
  \end{description}
\end{definition}
In other words \textsc{E-Maj-Sat} asks if there is an assignment to the first $k$ variables of $\Phi$ such that a majority of assignments to the remaining $d-k$ variables satisfies $\Phi$. 
There are three hurdles to take if we want to reduce this to \textsc{Relevant-Input}.
\begin{enumerate}
    \item Instead of assigning values to a given set of $k$ variables we can freely choose the set $S$ of size $k$.
    \item Instead of freely assigning values to a subset of variables we are given an assignment to all variables and can only chose to fix a subset of them and randomise the rest.
    \item Instead of checking whether the majority of assignments satisfies $\Phi$ we check if the fraction of satisfying assignments is larger or equal to some $\delta$.
\end{enumerate}
We address each of these hurdles and give a chain of polynomial-time reductions $\textsc{E-Maj-Sat}\preceq_p\textsc{IP1}\preceq_p\textsc{IP2}\preceq_p\textsc{Relevant-Input}$ in three steps with intermediate auxiliary problems \textsc{IP1} and \textsc{IP2}. The following observations will turn out to be useful.
\begin{remark} \label{rem:xor_prob}
    Let $\Phi$ and $\Psi$ be Boolean functions, not necessarily of different variables. Then
    \begin{align*}
    P(\Psi)=0 \quad&\Rightarrow\quad P\kl{\Phi \xor \Psi} = P(\Phi), \\
    P(\Psi)=1 \quad&\Rightarrow\quad P\kl{\Phi \xor \Psi} = 1-P(\Phi),
    \end{align*}
    and if $\Phi$ and $\Psi$ are independent, i.e. $P(\Phi \und \Psi) = P(\Phi)P(\Psi)$, also
    \begin{equation*}
    P(\Psi)=\frac{1}{2} \quad\Rightarrow\quad P\kl{\Phi \xor \Psi} = \frac{1}{2}.
    \end{equation*}
\end{remark}
\begin{lemma} \label{lem:change_uv}
  Let $B\colon\skl{0,1}^k\times\skl{0,1}^k\to\skl{0,1}$ and $\Psi\colon\skl{0,1}^k\times\skl{0,1}^k\times\skl{0,1}\to\skl{0,1}$ be defined as
  \begin{align*}
      B(\bfu,\bfv) &= \Und_{i=1}^k \lnot(u_i \xor v_i) \\
     \Psi(\bfu, \bfv, t) &= \kl{\Odr_{i=1}^k (u_i\xor v_i)} \und t.
  \end{align*}
  Then, for any $\Phi\colon\skl{0,1}^{k}\times\skl{0,1}^{d-k}\to\skl{0,1}$ and $A\colon\skl{0,1}^{k}\times\skl{0,1}^k\to\skl{0,1}$, we have
  \begin{equation*}
    P\kl{\Phi(\bfu,\bfr) \xor \Psi(\bfu,\bfv,t)\,\middle|\, A(\bfu,\bfv)} > \frac{1}{2} \quad \Longleftrightarrow \quad P\kl{\Phi(\bfu,\bfr) \,\middle|\, A(\bfu,\bfv), B(\bfu,\bfv)} > \frac{1}{2}.
  \end{equation*}
\end{lemma}
\begin{proof}
  We can rewrite $\Psi(\bfu,\bfv,t) = \kl{\lnot B(\bfu,\bfv)}\und t$ and therefore
  \begin{align*}
    P\kl{\Psi\,\middle|\,B} &= 0 \\
    P\kl{\Psi\,\middle|\, \neg B} &= \frac{1}{2}.
  \end{align*}
  Since $\Phi\,|\,A$ and $\Psi\,|\, A$ are conditionally independent given $\lnot B$ (in this case $\Psi$ depends on $t$ only), we obtain from \cref{rem:xor_prob} that
  \[
    P\kl{\Phi \xor \Psi\,\middle|\, A, B} = P\kl{\Phi\,\middle|\, A, B} 
  \]
  and
  \[
    P\kl{\Phi \xor \Psi\,\middle|\, A, \neg B} = \frac{1}{2}.
  \]
  Therefore
  \begin{align*}
    P\kl{\Phi \xor \Psi\,\middle|\, A} &= P\kl{\Phi\xor\Psi\,\middle|\, A, B}P\kl{B} + P\kl{\Phi\xor\Psi\,\middle|\, A, \lnot B}P\kl{\lnot B} \\
    &= P\kl{\Phi\,\middle|\, A, B}P\kl{B} + \frac{1}{2}\kl{1-P\kl{B}} \\
    &= \frac{1}{2} + \kl{P\kl{\Phi\,\middle|\, A, B}-\frac{1}{2}} P\kl{B}.
  \end{align*}
  This directly implies $P\kl{\Phi \xor \Psi\,\middle|\, A}>\frac{1}{2}$ if and only if $P\kl{\Phi\,\middle|\, A, B}>\frac{1}{2}$.
\end{proof}

Let us now come to the first step of the reductive chain. In this, we translate the option to freely assign the first $k$ variables into the choice of fixing a set of variables from a given assignment or randomising them. This choice is however still restricted to the first $k$ variables. 
\begin{definition}
  We define the \textsc{Intermediate Problem 1} (\textsc{IP1}) as follows.
  \begin{description}
    \item[Given:] $\Phi\colon\skl{0,1}^d\to\skl{0,1}$, $\bfx\in\skl{0,1}^d$ and $k\in\N$, $1\leq k\leq d$.
    \item[Decide:] Does there exist $S \subseteq\ekl{k}$ such that $P_{\bfy}\kl{\Phi(\bfy) \,\middle|\, \bfy_{S} = \bfx_{S}} > \frac{1}{2}$?
  \end{description}
\end{definition}
In other words \textsc{IP1} asks the questions if there is a subset of the first $k$ variables of $\Phi$, such that when fixing these to the values given by $\bfx$, a majority of assignments to the remaining variables satisfies $\Phi$. 
\begin{lemma}
$\textsc{E-Maj-Sat}\preceq_p\textsc{IP1}$, in particular \textsc{IP1} is $\SNP^{\SPP}$-hard.
\end{lemma}
\begin{proof}
  Let $\skl{\Phi, k}$ be an \textsc{E-Maj-Sat} instance. We will construct $\skl{\Phi^\prime,\bfx^\prime,k^\prime}$ that is a \emph{Yes}-instance for \textsc{IP1} if and only if $\skl{\Phi, k}$ is a \emph{Yes}-instance for \textsc{E-Maj-Sat}. For convenience we split the $d$ variables of $\Phi$ into the first $k$ variables and the remaining $d-k$ variables and denote this $\Phi(\bfx)=\Phi(\bfu,\bfr)$. The main idea is to duplicate the first $k$ variables and choose $\bfx^\prime$ in such a way that fixing the original variables or their duplicates corresponds to assigning zeros or ones in the \textsc{E-Maj-Sat} instance respectively. More precisely, we define
  \begin{itemize}
    \item $\Phi^\prime\colon\skl{0,1}^k\times\skl{0,1}^k\times\skl{0,1}^{d-k}\times\skl{0,1}\to\skl{0,1}$ as
    \[
    \Phi^\prime(\bfu, \bfv, \bfr, t)=\Phi(\bfu, \bfr)\xor \kl{ \Odr_{i=1}^k (u_i \xor v_i) \und t },
    \]
    \item $\mathbf{x}^\prime = \kl{\bfzero_k, \bfone_k, \bfzero_{d-k}, 0}\in\skl{0,1}^k\times\skl{0,1}^k\times\skl{0,1}^{d-k}\times\skl{0,1}$,
    \item $k^\prime = 2k$.
  \end{itemize}
  This is a polynomial time construction. With $\Psi$ as in \cref{lem:change_uv} we can rewrite $\Phi^\prime\kl{\bfu,\bfv,\bfr,t}=\Phi(\bfu,\bfr)\xor\Psi\kl{\bfu,\bfv,t}$.

  \paragraph{Necessity:}
  Assume that $\skl{\Phi, k}$ is a \emph{Yes}-instance for \textsc{E-Maj-Sat}. Then there exists an assignment $\bfu^\ast\in\skl{0,1}^k$ to the first $k$ variables of $\Phi$ such that  $P_{\bfr}\kl{\Phi(\bfu^\ast, \bfr)} > \frac{1}{2}$. Now choose $S^\prime=\skl{\,i\in\skl{1,\dots,k}\,:\,u^\ast_i=0\,}\cup\skl{\,i\in\skl{k+1,\dots,2k}\,:\,u^\ast_{i-k}=1\,}\subseteq\ekl{k^\prime}=\ekl{2k}$. Let $A\colon\skl{0,1}^{k}\times\skl{0,1}^k\to\skl{0,1}$ be given by
  \[ A(\bfu,\bfv) = \kl{\Und_{i\in S^\prime\cap\skl{1,\dots,k}} \lnot u_i} \und \kl{\Und_{i\in S^\prime\cap\skl{k+1,\dots,2k}} v_{i-k}} \] and $B$ as in \cref{lem:change_uv}. Note that $A$ depends on $S^\prime$ and thus implicitly on $\bfu^\ast$. In fact, $A(\bfu,\bfv)=1$ holds if and only if $ u_i^*=0 $ implies $ u_i=0 $ and $ u_i^*=1 $ implies $ v_i=1 $ for all $ i\in [k] $. In particular, we have $A(\bfu,\bfu)=1$ if an only if $\bfu=\bfu^\ast$. Also $B(\bfu,\bfv)=1$ if and only if $\bfu=\bfv$. Thus, by the choice of $A$, $B$, and $S^\prime$ we have
  \begin{align*}
    P_\bfr\kl{\Phi(\bfu^\ast,\bfr)} &= P\kl{\Phi(\bfu,\bfr)\,\middle|\, \bfu=\bfu^\ast} \\     
    &= P\kl{\Phi(\bfu,\bfr)\,\middle|\, A(\bfu,\bfu)} \\
    &= P\kl{\Phi(\bfu,\bfr)\,\middle|\, A(\bfu,\bfv), B(\bfu,\bfv)},
  \end{align*}
  and by the choice of $\bfx^\prime$, $A$, and $S^\prime$ we get
  \begin{align*}
    P_{\bfy^\prime}\kl{\Phi^\prime(\bfy^\prime) \,\middle|\, \bfy^\prime_{S^\prime} = \bfx^\prime_{S^\prime}} &= P\kl{\Phi^\prime(\bfu,\bfv,\bfr,t)\,\middle|\, A(\bfu,\bfv)} \\
    &= P\kl{\Phi(\bfu,\bfr)\xor\Psi\kl{\bfu,\bfv,t}\,\middle|\, A(\bfu,\bfv)}.
  \end{align*}
  Since $P_{\bfr}\kl{\Phi(\bfu^\ast, \bfr)} > \frac{1}{2}$, we can use \cref{lem:change_uv} to conclude $P_{\bfy^\prime}\kl{\Phi^\prime(\bfy^\prime) \,\middle|\, \bfy^\prime_{S^\prime} = \bfx^\prime_{S^\prime}}>\frac{1}{2}$ which shows that $\skl{\Phi^\prime, \bfx^\prime, k^\prime}$ is a \emph{Yes}-instance for \textsc{IP1}.
  
  \paragraph{Sufficiency:}
  Now, conversely, assume that $\skl{\Phi^\prime,\bfx^\prime,k^\prime}$ is a \emph{Yes}-instance for \textsc{IP1}. Then there exists $S^\prime\subseteq\ekl{k^\prime}=\ekl{2k}$ such that $P_{\bfy^\prime}\kl{\Phi^\prime(\bfy^\prime) \,\middle|\, \bfy^\prime_{S^\prime} = \bfx^\prime_{S^\prime}}>\frac{1}{2}$. Following the same grouping of variables as before, we write $\bfx^\prime=\kl{\bfu^\prime,\bfv^\prime,\bfr^\prime,t^\prime}$ and observe that $u^\prime_i\neq v^\prime_i$ for all $i\in\ekl{k}$. Towards a contradiction, assume there exists an $i\in\ekl{k}$ with $i\in S^\prime$ and $i+k\in S^\prime$. Then $ u_i\ne v_i $ if $ (\bfu,\bfv)_{S^\prime} = (\bfu^\prime,\bfv^\prime)_{S^\prime}$ and hence
  \begin{equation*}
    P\kl{\Psi(\bfu,\bfv,t)\,\middle|\, (\bfu,\bfv)_{S^\prime} = (\bfu^\prime,\bfv^\prime)_{S^\prime}} = P\kl{\Psi(\bfu,\bfv,t)\,\middle|\, \lnot B(\bfu,\bfv)} = \frac{1}{2}.
  \end{equation*}
  Thus, \cref{rem:xor_prob} would imply $P_{\bfy^\prime}\kl{\Phi^\prime(\bfy^\prime)\,\middle|\, \bfy^\prime_{S^\prime} = \bfx^\prime_{S^\prime}}=\frac{1}{2}$, which is a contradiction. Therefore for all $i\in\ekl{k}$ we can have $i\in S^\prime$ or $i+k\in S^\prime$ but not both. Similarly, if we assume there exists an $i\in\ekl{k}$ with neither $i\in S^\prime$ nor $i+k\in S^\prime$ we have
  \begin{equation*}
    P_{\bfy^\prime}\kl{\Phi^\prime(\bfy^\prime) \,\middle|\, \bfy^\prime_{S^\prime} = \bfx^\prime_{S^\prime}} = P_{\bfy^\prime}\kl{\Phi^\prime(\bfy^\prime) \,\middle|\, \bfy^\prime_{S^\prime \cup \skl{i+k}} = \bfx^\prime_{S^\prime \cup \skl{i+k}}}.
  \end{equation*}
  This holds because the only difference is that the clause $u_i\xor v_i$ in $\Psi$ is replaced by $u_i\xor 1=\lnot u_i$ and this does not change the overall probability. So without loss of generality we can assume that for each $i\in\ekl{k}$ exactly one of the cases $i\in S^\prime$ or $i+k\in S^\prime$ occurs. 
  Then we can define $\bfu^\ast\in\skl{0,1}^k$ as $u_i^\ast=0$ if $i\in S^\prime$ and $u_i^\ast=1$ otherwise. We observe that $S^\prime$ and $\bfu^\ast$ are exactly as in the previous step and the rest of the proof follows analogously. Again we use \cref{lem:change_uv} and conclude from $P_{\bfy^\prime}\kl{\Phi^\prime(\bfy^\prime) \,\middle|\, \bfy^\prime_{S^\prime} = \bfx^\prime_{S^\prime}}>\frac{1}{2}$ that $P_{\bfr}\kl{\Phi(\bfu^\ast, \bfr)} > \frac{1}{2}$. This shows that $\skl{\Phi, k}$ is a \emph{Yes}-instance for \textsc{E-Maj-Sat}.
\end{proof}

We continue with the second step of the reductive chain. We translate the question whether a majority of assignments to the non-fixed variables satisfies a Boolean formula into the question whether the fraction of assignments leaving the function value unchanged is larger or equal to some $\delta$.
\begin{definition}
  For $\delta\in (0,1]$ we define the \textsc{Intermediate Problem 2} (\textsc{IP2}) as follows.
  \begin{description}
    \item[Given:] $\Phi\colon\skl{0,1}^d\to\skl{0,1}$, $\bfx\in\skl{0,1}^d$ and $k\in\N$, $1\leq k\leq d$.
    \item[Decide:] Does there exist $S \subseteq\ekl{k}$ such that $P_{\bfy}\kl{\Phi(\bfy)\,\middle|\, \bfy_{S} = \bfx_{S}} \geq \delta$?
  \end{description}
\end{definition}
In other words \textsc{IP2} asks the question if there exists a subset of the first $k$ variables of $\Phi$ such that when fixing these to the values given by $\bfx$ a fraction $\delta$ of assignments to the remaining variables leaves $\Phi$ unchanged. In order to relate this to \textsc{IP1}, we have to change the acceptance probability from $\frac{1}{2}$ to $\delta$. This can be done using the following \namecref{lem:prob_transform}.
\begin{lemma} \label{lem:prob_transform}
  Given $ d\in\N $ and $0<\delta_1<\delta_2<1$, there exists a monotone function $\Pi\colon\skl{0,1}^n\to\skl{0,1}$ such that $\Pi(\bfzero_n)=0$, $\Pi(\bfone_n)=1$, and for all $\Phi\colon \skl{0,1}^d \to \skl{0,1}$ we have 
  \begin{equation*}
    P_\bfy(\Phi(\bfy)) > \delta_1 \quad\Longleftrightarrow\quad P_{(\bfy,\bfr)}(\Phi(\bfy) \odr \Pi(\bfr)) \geq \delta_2
  \end{equation*}
  with $n\in\mathcal{O}\kl{\kl{d+\log_2\kl{\frac{1-\delta_1}{1-\delta_2}}}^2+\log_2\kl{\frac{1}{\delta_2-\delta_1}}}$. It can be constructed in $\mathcal{O}\kl{n}$ time.
\end{lemma}
The constructive proof of \cref{lem:prob_transform} can be found in \cref{apx:prob_transform}. 
\begin{lemma}
For $\delta\in\left[\frac{1}{2},1\right)$ we have $\textsc{IP1}\preceq_p\textsc{IP2}$, in particular in this case \textsc{IP2} is $\SNP^{\SPP}$-hard.
\end{lemma}

\begin{proof}
  Let $\skl{\Phi,\bfx,k}$ be an \textsc{IP1} instance. We will construct $\skl{\Phi^\prime,\bfx^\prime,k^\prime}$ that is a \emph{Yes}-instance for \textsc{IP2} if and only if $\skl{\Phi,\bfx,k}$ is a \emph{Yes}-instance for \textsc{IP1}. Let $\Pi\colon\skl{0,1}^n\to\skl{0,1}$ be as in \cref{lem:prob_transform} for $\delta_1=\frac{1}{4}$ and $\delta_2=\delta$. We define
  \begin{itemize}
    \item  $\Phi^\prime\colon\skl{0,1}^d\times\skl{0,1}\times\skl{0,1}^n\to\skl{0,1}\colon (\bfy,t,\bfr) \mapsto \kl{\Phi(\bfy)\und t}\odr \Pi(\bfr)$,
    \item $\bfx^\prime = \kl{\bfx, 1, \bfone_n}\in\skl{0,1}^d\times\skl{0,1}\times\skl{0,1}^n$,
    \item $k^\prime = k$.
  \end{itemize}
  This is a polynomial time construction. By the choice of $\Phi^\prime$ and $\bfx^\prime$ we guarantee $\Phi^\prime(\bfx^\prime)=1$ regardless of the value of $\Phi(\bfx)$ since $ \Pi(\bfone_n)=1 $.\\
  Then, for all $S^\prime=S\subseteq [k]=\ekl{k^\prime}$, we have
  \begin{alignat*}{2}
    &&P_\bfy\kl{\Phi(\bfy) \,\middle|\, \bfy_S = \bfx_S}&>\frac{1}{2} \\
    &\Longleftrightarrow\quad & P_{(\bfy,t)}\kl{\Phi(\bfy) \und t \,\middle|\, \bfy_S = \bfx_S} &> \frac{1}{4}\\
    &\Longleftrightarrow\quad & P_{(\bfy,t,\bfr)}\kl{(\Phi(\bfy) \und t)\odr\Pi(\bfr) \,\middle|\, \bfy_S = \bfx_S} &\geq \delta \\
&\Longleftrightarrow\quad & P_{\bfy^\prime}\kl{\Phi^\prime(\bfy^\prime) = \Phi^\prime(\bfx^\prime) \,\middle|\, \bfy^\prime_{S^\prime} = \bfx^\prime_{S^\prime}} &\geq \delta.
  \end{alignat*}
  Thus $\skl{\Phi,\bfx,k}$ is a \emph{Yes}-instance for \textsc{IP1} if and only if $\skl{\Phi^\prime,\bfx^\prime,k^\prime}$ is a \emph{Yes}-instance for \textsc{IP2}.
\end{proof}

We continue with the third and final step of the reductive chain. We translate the option to choose which of the first $k$ variables to fix into the choice of fixing any set of at $k$ variables.

\begin{lemma}
For $\delta\in\left[\frac{1}{2},1\right)$ we have $\textsc{IP2}\preceq_p\textsc{Relevant-Input}$, in particular in this case \textsc{Relevant-Input} is $\SNP^{\SPP}$-hard.
\end{lemma}

\begin{proof}
  Let $\skl{\Phi,\bfx,k}$ be an \textsc{IP2} instance. We will construct $\skl{\Phi^\prime,\bfx^\prime,k^\prime}$ that is a \emph{Yes}-instance for \textsc{Relevant-Input} if and only if $\skl{\Phi,\bfx,k}$ is a \emph{Yes}-instance for \textsc{IP2}. For convenience, we split the $d$ variables of $\Phi$ into the first $k$ variables and the remaining $d-k$ variables and denote this $\Phi(\bfx)=\Phi(\bfu,\bfr)$. The main idea is to extend $\Phi$ with clauses that force the set $S$ to be chosen from the first $k$ variables. More precisely, we define
  \begin{itemize}
    \item  $\Phi^\prime\colon\skl{0,1}^k\times\skl{0,1}^k\times\skl{0,1}^{d-k}\times\skl{0,1}^{d-k}\times\skl{0,1}^{d-k}\to\{0,1\}$ with 
    \[
    \Phi^\prime(\bfu,\bfv,\bfr_1,\bfr_2,\bfr_3) = \kl{\Phi(\bfu,\bfr_1\xor\bfr_2\xor\bfr_3)\xor(\lnot\Phi(\bfx))}\und \kl{ \Und_{i=1}^k \kl{ \kl{u_i\xor \lnot x_i} \odr v_i}},
    \]
    where $\bfr_1\xor\bfr_2\xor\bfr_3$ is understood component-wise,
    \item $\bfx^\prime = \kl{\bfx_{[k]}, \bfone_k, \bfx_{[k]^c}, \bfx_{[k]^c}, \bfx_{[k]^c}}\in\skl{0,1}^k\times\skl{0,1}^k\times\skl{0,1}^{d-k}\times\skl{0,1}^{d-k}\times\skl{0,1}^{d-k}$,
    \item $k^\prime = k$.
  \end{itemize}
   This is a polynomial time construction. By the choice of $\Phi^\prime$ and $\bfx^\prime$ we guarantee $\Phi^\prime(\bfx^\prime)=1$ regardless of the value of $\Phi(\bfx)$.
  
  \paragraph{Necessity:}
  Assume that $\skl{\Phi,\bfx,k}$ is a \emph{Yes}-instance for \textsc{IP2}. Then there exists $S\subseteq\ekl{k}$ such that $P_\bfy\kl{\Phi(\bfy) = \Phi(\bfx) \,|\, \bfy_S = \bfx_S )}\geq\delta$. Now choose 
  \[
  S^\prime = S\cup\skl{\,i\in\skl{k+1,\dots,2k}\,:\,i-k\notin S\,}.
  \]
  Then $|S^\prime|=|S|+(k-|S|) = k = k^\prime$ and for each $i\in\ekl{k}$ exactly one of the cases $i\in S^\prime$ or $i+k\in S^\prime$ can occur. The former corresponds to fixing $u_i=x_i^\prime=x_i$ and the latter to fixing $v_i=1$. Therefore
  \[P_{(\bfu,\bfv)}\kl{\Und_{i=1}^k \kl{ \kl{u_i\xor \lnot x_i} \odr v_i} \,\middle|\, \kl{\bfu,\bfv}_{S^\prime} = \bfx^\prime_{S^\prime}} = 1,\]
  which means, conditioned on $\kl{\bfu,\bfv}_{S^\prime} = \bfx^\prime_{S^\prime}$, the probability of satisfying $\Phi^\prime$ only depends on $\Phi(\bfu,\bfr_1\xor\bfr_2\xor\bfr_3)$. Now, since the random vector $\bfr_1\xor\bfr_2\xor\bfr_3$ is independent of this condition, it has the exact same distribution as the random vector $\bfr$ and we obtain
  \begin{align*}
    P_{\bfy^\prime}\kl{\Phi^\prime(\bfy^\prime)=\Phi^\prime(\bfx^\prime) \,\middle|\, \bfy^\prime_{S^\prime} = \bfx^\prime_{S^\prime}} &= P_{\bfy^\prime}\kl{\Phi^\prime(\bfy) \,\middle|\, \bfy^\prime_{S^\prime} = \bfx^\prime_{S^\prime}} \\
    &= P\kl{\Phi^\prime(\bfu,\bfv,\bfr_1,\bfr_2,\bfr_3)\,\middle|\, \kl{\bfu,\bfv}_{S^\prime}=\bfx^\prime_{S^\prime} } \\
    &= P\kl{\Phi^\prime(\bfu,\bfv,\bfr_1,\bfr_2,\bfr_3)\,\middle|\, \bfu_S=\bfx_S, \bfv_{S^c} =\bfone } \\ 
    &= P\kl{\Phi(\bfu,\bfr)\xor(\lnot\Phi(\bfx))\,\middle|\, \bfu_S=\bfx_S} \\
    &= P\kl{\Phi(\bfu,\bfr) = \Phi(\bfx)\,\middle|\, \bfu_S=\bfx_S} \\
    &= P_\bfy\kl{\Phi(\bfy)=\Phi(\bfx) \,\middle|\, \bfy_{S} = \bfx_{S}}\geq\delta.
  \end{align*}
  This shows that $S^\prime$ is a $\delta$-relevant set for $\Phi^\prime$ and $\bfx^\prime$. Therefore $\skl{\Phi^\prime,\bfx^\prime,k^\prime}$ is a \emph{Yes}-instance for \textsc{Relevant-Input}.
  
  \paragraph{Sufficiency:}
  Now, conversely, assume that $\skl{\Phi^\prime,\bfx^\prime,k^\prime}$ is a \emph{Yes}-instance for \textsc{Relevant-Input}. Then there exists a $\delta$-relevant set $S^\prime\subseteq\ekl{2k+3(d-k)}$ for $\Phi^\prime$ and $\bfx^\prime$ with $|S^\prime|\leq k^\prime=k$. Hence, we have
  \[
  P_{\bfy^\prime}\kl{\Phi^\prime(\bfy^\prime) \,\middle|\, \bfy^\prime_{S^\prime} = \bfx^\prime_{S^\prime}} =  P_{\bfy^\prime}\kl{\Phi^\prime(\bfy^\prime)=\Phi^\prime(\bfx^\prime) \,\middle|\, \bfy^\prime_{S^\prime} = \bfx^\prime_{S^\prime}} \geq\delta.
  \]
  Without any conditions on $\bfu$ and $\bfv$ we have $P\kl{(u_i\xor \lnot x_i) \odr v_i}=\frac{3}{4}$ for all $i\in\ekl{k}$. Since $\kl{\frac{3}{4}}^3 < \frac{1}{2}\leq\delta$, we have $i\in S^\prime$ or $i+k\in S^\prime$ for all $i\in\ekl{k}$ except for at most two cases. Hence, $|S^\prime|\leq k$ implies $|S^\prime\cap \ekl{2k}^c|\leq 2$. Therefore, $\bfr_1\xor\bfr_2\xor\bfr_3$ conditioned on $\bfy^\prime_{S^\prime} = \bfx^\prime_{S^\prime}$ has the same distribution as $\bfr_1\xor\bfr_2\xor\bfr_3$ without the condition. So, without loss of generality, we can even assume $S^\prime\cap\ekl{2k}^c=\emptyset$. 
  
  Similarly, if $i\in S^\prime$ we have $P\kl{(u_i\xor \lnot x_i)\odr v_i\,\middle|\, \kl{\bfu,\bfv}_{S^\prime} = \bfx^\prime_{S^\prime}} = 1$ and additionally having $i+k\in S^\prime$ could not increase the probability of satisfying $\Phi^\prime$. Contrary, if $i\notin S^\prime$, we have 
  \[
  P\kl{(u_i\xor \lnot x_i)\odr v_i\,\middle|\, \kl{\bfu,\bfv}_{S^\prime} = \bfx^\prime_{S^\prime}} = \frac{1}{2}+\frac{1}{2}P\kl{v_i\,\middle|\, \kl{\bfu,\bfv}_{S^\prime} = \bfx^\prime_{S^\prime}},
  \]
  which is one if $i+k\in S^\prime$ and $\frac{3}{4}$ otherwise. So including $i+k$ in $S^\prime$ does not decrease the probability. Altogether, without loss of generality we can assume $S^\prime\subseteq [2k]$, $|S^\prime|=k$ and for each $i\in\ekl{k}$ exactly one of the cases $i\in S^\prime$ or $i+k\in S^\prime$ occurs. 
  
  We now choose $S=S^\prime\cap\ekl{k}$. Then the rest of the proof proceeds exactly as above and we conclude
  \begin{align*}
    P_\bfy\kl{\Phi(\bfy)=\Phi(\bfx) \,\middle|\, \bfy_{S} = \bfx_{S}} &= P\kl{\Phi(\bfu,\bfr)\xor(\lnot\Phi(\bfx))\,\middle|\, \bfu_S=\bfx_S} \\
    &= P\kl{\Phi^\prime(\bfu,\bfv,\bfr_1,\bfr_2,\bfr_3)\,\middle|\, \bfu_S=\bfx_S, \bfv_{S^c} =\bfone } \\ 
    &= P\kl{\Phi^\prime(\bfu,\bfv,\bfr_1,\bfr_2,\bfr_3)\,\middle|\, \kl{\bfu,\bfv}_{S^\prime}=\bfx^\prime_{S^\prime} } \\
    &=  P_{\bfy^\prime}\kl{\Phi^\prime(\bfy^\prime_{S^\prime}) \,\middle|\, \bfy^\prime_{S^\prime} = \bfx^\prime_{S^\prime}} \\
    &=  P_{\bfy^\prime}\kl{\Phi^\prime(\bfy^\prime) = \Phi^\prime(\bfx^\prime) \,\middle|\, \bfy^\prime_{S^\prime} = \bfx^\prime_{S^\prime}} \geq \delta,
  \end{align*}
  which shows that $\skl{\Phi,\bfx,k}$ is a \emph{Yes}-instance for \textsc{IP2}. 
\end{proof}

Altogether, this section finishes the first part of the proof of \cref{thm:npppcomplete}.

\subsection{\textsc{Relevant-Input} is contained in \texorpdfstring{$\SNP^\SPP$}{NP\^{}PP}}\label{sec:npppcontained}
We now come to the second part of the proof of \cref{thm:npppcomplete}. We will show that \textsc{Relevant-Input} is in fact contained in $\SNP^\SPP$ meaning that it can be solved in polynomial time by a non-deterministic Turing machine with access to a $\SPP$-oracle. The following \namecref{lem:prob_transform2} will be useful.

\begin{lemma} \label{lem:prob_transform2}
  Given $ d\in\N $ and $0<\delta_1<\delta_2<1$, there exists a monotone function $\Pi\colon\skl{0,1}^n\to\skl{0,1}$ such that, for all $\Phi\colon \skl{0,1}^d \to \skl{0,1}$, we have
  \begin{equation*}
    P_\bfy(\Phi(\bfy)) \geq \delta_2 \quad\Longleftrightarrow\quad P_{(\bfy,\bfr)}(\Phi(\bfy) \und \Pi(\bfr)) > \delta_1
  \end{equation*}
  with $n\in\CO\kl{\kl{d+\log_2\kl{\frac{\delta_2}{\delta_1}}}^2}$. It can be constructed in $\mathcal{O}\kl{n}$ time.
\end{lemma}
The constructive proof of \cref{lem:prob_transform2} can be found in \cref{apx:prob_transform2}. 

\begin{lemma} \label{lem:isinnppp}
For $\delta\in \left(0,1\right)$ the \textsc{Relevant-Input} problem is contained in $\SNP^{\SPP}$.
\end{lemma}

We will prove this for $\delta\in \left(\frac{1}{2},1\right)$ by lowering the probability threshold from $\delta$ to $\frac{1}{2}$. The case $\delta\in \left(0, \frac{1}{2}\right]$ can be treated analogously by raising the threshold.  

\begin{proof}
Let $\skl{\Phi,\bfx,k}$ be an instance of \textsc{Relevant-Input}. It suffices to show that the decision problem whether a given set $S\subseteq\ekl{d}$ is $\delta$-relevant for $\Phi$ and $\bfx$ is in $\SPP$. Without loss of generality we can assume $\Phi(\bfx)=1$. Otherwise we could consider $\lnot \Phi$ instead. Now, choose $\Pi\colon\skl{0,1}^n\to\skl{0,1}$ as in \cref{lem:prob_transform2} for $\delta_1=\frac{1}{2}$ and $\delta_2=\delta$. Then
\begin{equation*}
    P_\bfy\kl{\Phi(\bfy)\,\middle|\, \bfy_S=\bfx_S} \geq \delta \quad\Longleftrightarrow\quad P_{(\bfy,\bfr)}\kl{\Phi(\bfy)\und\Pi(\bfr)\,\middle|\,\bfy_S=\bfx_S} > \frac{1}{2}.
\end{equation*}
A probabilistic Turing machine can now draw a random assignment $(\bfy,\bfr)$ conditioned on $\bfy_S=\bfx_S$ and evaluate $\Phi(\bfy)\und\Pi(\bfr)$. Thus, the machine will answer \emph{Yes} with probability strictly greater than $\frac{1}{2}$ if and only if $S$ is $\delta$-relevant. This means the subproblem of checking a set for $\delta$-relevance is contained in $\SPP$.

A non-deterministic Turing-machine  with a $\SPP$-oracle can thus search over all possible sets $S\subseteq\ekl{d}$ with $\bkl{S} \leq k$ and check whether any of them is $\delta$-relevant using the oracle.
\end{proof}

\section{Variations of the Problem Formulation}\label{sec:probvariants}
We want to consider two variations of the \textsc{Relevant-Input} problem. The first variation relaxes the requirement to check if a candidate set $S$ is exactly $\delta$-relevant or not by introducing a probability gap $\gamma$. In short, we then ask if a $\delta$-relevant set of size $k$ exists or if all sets of size $k$ are not even $(\delta-\gamma)$-relevant. 

The second variation concerns the optimisation version of the problem. Here we introduce a set size gap and relax the requirement to find the smallest $\delta$-relevant set. Instead for $k<m$ we ask if a $\delta$-relevant set of size $k$ exists or if all relevant sets must be of size at least $m$.

We show that these problems remain hard to solve (even in combination, that is with both a gap in probability and set size). This can be used to show that no polynomial time approximation algorithm for \textsc{Min-Relevant-Input} with approximation factor better than the trivial factor $d$ can exists unless $\SP=\SNP$.
Due to the connection between Boolean functions and neural networks, as described in \cref{sec:introduction}, this inapproximability result shows theoretical limitations of interpretation methods for neural network decision.

\subsection{The Probability Gap}\label{sec:gapped}
Probabilities and expectation values may be hard to calculate in theory, yet are often easy to approximate in practice. Checking whether a Boolean function is satisfied by more than a fraction $\delta$ of the possible assignments to its variables can often be done probabilistically. This only fails if the true probability is close to $\delta$. These edge cases cause the hardness of the problem. Yet, in our scenario we do not necessarily care about their resolution. In fact, the choice of $\delta$ in the definition of $\delta$-relevant sets is somewhat arbitrary. Thus, it seems impractical to defend the hardness of the \textsc{Relevant-Input} problem with the exact evaluation of probabilities.

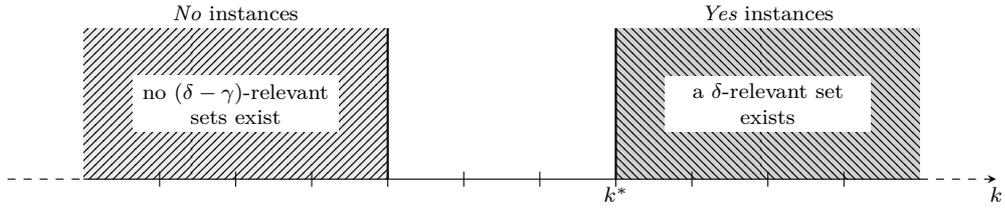
\begin{figure}
    \centering
    \scriptsize
\begin{tikzpicture}

\fill[gray!15] (-1,0) rectangle (3,2);
\fill[pattern=north east lines] (-1,0) rectangle (3,2);
\draw[thick] (3,0) -- (3,2);
\node[rectangle, fill=white] at (1,1) {\parbox{2.5cm}{\centering no $(\delta-\gamma)$-relevant sets exist}};
\node[above] at (1,2) {\emph{No} instances};

\fill[gray!35] (6,0) rectangle (10,2);
\fill[pattern=north west lines] (6,0) rectangle (10,2);
\draw[thick] (6,0) -- (6,2);
\node[rectangle, fill=white] at (8,1) {\parbox{2.5cm}{\centering a $\delta$-relevant set exists}};
\node[above] at (8,2) {\emph{Yes} instances};

\node[below] at (6,0) {$k^\ast$};

\draw (-1,0) -- (10,0);
\draw[dashed] (-2,0) -- (-1,0);
\draw[dashed, ->, >=stealth] (10,0) -- (11,0) node[below] {$k$};
\foreach \x in {0, 1, 2, 3, 4, 5, 6, 7, 8, 9}
    \draw (\x, -0.1) -- (\x, 0.1);
    
\end{tikzpicture}
    \caption{Visualisation of the \textsc{Gapped-Relevant-Input} problem for some fixed $\Phi$ and $\bfx$ and for various $k$. In the unmarked region in the centre no $\delta$-relevant set exists but $\widetilde{\delta}$-relevant sets could exist for any $\widetilde{\delta}<\delta$, in particular also for $\widetilde{\delta}=\delta-\gamma$. In this region we do not expect an answer for the gapped problem. The solution $k^\ast$ of the ungapped optimisation problem \textsc{Min-Relevant-Input} is the left boundary of the \emph{Yes}-instance region.}
    \label{fig:gapped_problem}
\end{figure}

Therefore, we introduce a variant of the problem including a probability gap. This can be seen as a promise problem with the promise that all sets $S$ are either $\delta$-relevant or not even $(\delta-\gamma)$-relevant. Alternatively, this can be seen as the \textsc{Relevant-Input} problem where we want to answer \emph{Yes} if a $\delta$-relevant set of size $k$ exists but only want to answer \emph{No} if all sets of size $k$ are not even $(\delta-\gamma)$-relevant. For cases inbetween we do not expect an answer at all or do not care about the exact answer. This is illustrated in \cref{fig:gapped_problem}.

\begin{definition}
For $\delta \in (0,1]$ and $\gamma \in [0,\delta)$ we define the \textsc{Gapped-Relevant-Input} problem as follows.
\begin{description}
  \item[Given:] $\Phi\colon\skl{0,1}^d\to\skl{0,1}$, $\bfx\in\skl{0,1}^d$, and $k\in\N$, $1\leq k\leq d$.
  \item[Decide:]~
    \begin{itemize}
      \item[\emph{Yes}:] There exists $S \subseteq[d]$ with $\bkl{S}\leq k$ and $S$ is $\delta$-relevant for $\Phi$ and $\bfx$.
      \item[\emph{No}:] All $S \subseteq [d]$ with $\bkl{S}\leq k$ are not $(\delta-\gamma)$-relevant for $\Phi$ and $\bfx$.
    \end{itemize}
\end{description}
\end{definition}
For $ \gamma=0 $ we exactly retrieve the original \textsc{Relevant-Input} problem but for $\gamma>0$ this is an easier question.

\begin{lemma}
For $\delta\in\left(0,1\right)$ and $\gamma\in (0,\delta)$ the \textsc{Gapped-Relevant-Input} problem is contained 
in $\SNP^{\SBPP}$.
\end{lemma}

\begin{proof}
    
    Let $\{\Phi,\bfx,k\}$ be an instance of \textsc{Gapped-Relevant-Input}. It suffices to show that the decision problem whether a given set $S\subseteq\ekl{d}$ is either $\delta$-relevant (\emph{Yes}) or not $(\delta-\gamma)$-relevant (\emph{No}) for $\Phi$ and $\bfx$ is in $\SBPP$. To see this, we describe an explicit algorithm with bounded error probability.
    
    Draw $n = \ceil{\frac{2\ln(3)}{\gamma^2}}$ independent random binary vectors $\bfb^{(i)} \in \skl{0,1}^{d-\bkl{S}}$ for $i \in [n]$ from the uniform distribution on $\skl{0,1}^{d-\bkl{S}}$ and define $\bfy^{(i)} \in \skl{0,1}^{d}$ as $\bfy^{(i)}_S = \bfx_S$ and $\bfy^{(i)}_{S^c}= \bfb^{(i)}$. Set
    \[
        \xi = \frac{1}{n}\sum_{i=1}^n \xi_i \quad\text{where}\quad  \xi_i = 
        \begin{cases}
        1,\quad &\text{if }\Phi(\bfx)=\Phi\kl{\bfy^{(i)}} \\[.75em]
        0,\quad &\text{if }\Phi(\bfx) \neq \Phi\kl{\bfy^{(i)}}
        \end{cases}
        \quad\text{for}\quad i=1,\dots,n.
    \]
    Then answer \emph{No} if $\xi < \delta - \frac{\gamma}{2}$ and answer \emph{Yes} if $\xi \geq \delta - \frac{\gamma}{2}$.
    
    The random variables $\xi_i$ are independently and identically Bernoulli distributed variables with 
    \[
    p=\E\ekl{\xi_i}=\E\ekl{\xi} = P_\bfy\kl{\Phi(\bfy_S) = \Phi(\bfx) \,|\, \bfy_S=\bfx_S}.
    \]
    Therefore $S$ is $\delta$-relevant if $p\geq \delta$ and not $(\delta-\gamma)$-relevant if $p<\delta-\gamma$. We use Hoeffding's inequality \cite{hoeffding1994probability} to bound the error probability of the algorithm. Firstly, assume $p\geq\delta$. Then we make an error if $\xi < \delta - \frac{\gamma}{2}$, which implies $p-\xi>\frac{\gamma}{2}$. The probability for this event can be bounded by
    \[
        P\kl{p-\xi > \frac{\gamma}{2}} \leq e^{- \frac{n\gamma^2}{2}} \leq \frac{1}{3}.
    \]
    Secondly, assume $p<\delta-\gamma$. Then we can bound the probability that $\xi\geq \delta-\frac{\gamma}{2}$, and thus $\xi-p>\frac{\gamma}{2}$, by
    \[
        P\kl{\xi -p > \frac{\gamma}{2}} \leq e^{- \frac{n\gamma^2}{2}} \leq \frac{1}{3}.
    \]
    Altogether the algorithm answers correctly with probability $\frac{2}{3}$ showing that the problem lies in $\SBPP$.
    
    A non-deterministic Touring machine  with $\SBPP$-oracle can thus search over all possible sets $S\subseteq\ekl{d}$ with $\bkl{S} \leq k$ and check if any of them is $\delta$-relevant or all of them are not $(\delta-\gamma)$-relevant.
\end{proof}

Similar to the original problem formulation we can also state an optimisation version of the gapped problem. In this case, we relax the optimality condition on the set size $k$ by allowing also sizes in the region between \emph{Yes}- and \emph{No}-instances of \textsc{Gapped-Relevant-Input} (cf.\ \cref{fig:gapped_problem}). In other words, we want to find any $k$ that is large enough so that it is not a \emph{No}-instance for the gapped problem but not larger than the optimal solution of the ungapped minimisation problem. Strictly speaking, this results in a search problem and not an optimisation problem. However, problems of this type can be referred to as weak optimisation problems \cite{GroetschelLovaszSchrijver1988optimization}.

\begin{definition} \label{def:min_gap}
For $\delta\in(0,1]$ and $\gamma\in[0,\delta)$ we define the \textsc{Min-Gapped-Relevant-Input} problem as follows.
\begin{description}
  \item[Given:] $\Phi\colon\skl{0,1}^d\to\skl{0,1}$ and $\bfx\in\skl{0,1}^d$.
  \item[Find:] $k\in\N$, $1\leq k \leq d$ such that
  \begin{enumerate}[label=(\roman*)]
    \item There exists $S \subseteq [d]$ with $\bkl{S}=k$ and $S$ is $(\delta-\gamma)$-relevant for $\Phi$ and $\bfx$.
    \item All $S \subseteq [d]$ with $\bkl{S}<k$ are not $\delta$-relevant for $\Phi$ and $\bfx$.
\end{enumerate}
\end{description}
\end{definition}

Note that both for \textsc{Gapped-Relevant-Input} and \textsc{Min-Gapped-Relevant-Input} a solution for $\gamma_1$ will always also be a solution for $\gamma_2>\gamma_1$. Specifically, being able to solve the ungapped problems introduced in \cref{sec:problem} provides a solution to the gapped problems for any $\gamma>0$.

\subsection{The Set Size Gap (Approximability)}\label{sec:inapproximability}

Even the gapped version of the minimisation problem is hard to approximate. We prove this by introducing another intermediate problem which we show to be $\SNP$-hard but which would be in $\SP$ if there exists a ``good'' polynomial time approximation algorithm for \textsc{Min-Gapped-Relevant-Input}. As already mentioned before, strictly speaking \textsc{Min-Gapped-Relevant-Input} is not an optimisation but a search problem. In order to give a meaning to the concept of approximation factors we use the following convention.
\begin{definition}
An algorithm for \textsc{Min-Gapped-Relevant-Input} has an approximation factor $c\geq 1$ if, for any instance $\skl{\Phi,\bfx}$, it produces an approximate solution $k$ such that there exists a true solution $\widetilde{k}$ (satisfying both conditions in \cref{def:min_gap}) with $\widetilde{k}\le k\leq c\widetilde{k}$.
\end{definition}

\begin{figure}
    \centering
    \scriptsize
\begin{tikzpicture}






\fill[gray!5] (-2,0) -- (-2,13) -- (11,13) -- cycle;

\fill[gray!35] (6,8) -- (10,12) -- (6,12) -- cycle;
\fill[pattern=north west lines] (6,8) -- (10,12) -- (6,12) -- cycle;

\fill[gray!15] (-1,1) -- (3,5) -- (-1,5) -- cycle;
\fill[pattern=north east lines] (-1,1) -- (3,5) -- (-1,5) -- cycle;

\node[rectangle, fill=white] at (0.25,4) {\emph{No}-instances};
\node[rectangle, fill=white] at (7.25,11) {\emph{Yes}-instances};

\draw[thick] (6,0) -- (6,13);
\draw[thick,->] (6,1) -- (6.3,1) node[right] {\parbox{2.5cm}{\centering a $\delta$-relevant set of size $k$ exists}};
\draw[thick] (3,0) -- (3,13); 
\draw[thick,->] (3,1) -- (2.7,1) node[left] {\parbox{2.5cm}{\centering no $(\delta-\gamma)$-relevant set of size $k$ exists}};

\draw[thick] (-2,5) -- (11,5);
\draw[thick,->] (9,5) -- (9,4.7) node[below] {\parbox{2.5cm}{\centering no $(\delta-\gamma)$-relevant set of size $m$ exists}};
\draw[thick] (-2,8) -- (11,8);
\draw[thick,->] (9,8) -- (9,8.3) node[above] {\parbox{2.5cm}{\centering a $\delta$-relevant set of size $m$ exists}};

\draw[dashed] (-2,0) -- (-1,1);
\draw (-1,1) -- (10,12);
\draw[dashed] (10,12) -- (11,13) node[right]{$k=m$};

\draw (-1,0) -- (10,0);
\draw[dashed] (-2,0) -- (-1,0);
\draw[dashed, ->, >=stealth] (10,0) -- (11,0) node[below] {$k$};
\foreach \x in {0, 1, 2, 3, 4, 5, 6, 7, 8, 9}
    \draw (\x, -0.1) -- (\x, 0.1);
    
\draw (-2,1) -- (-2,12);
\draw[dashed] (-2,0) -- (-2,1);
\draw[dashed, ->, >=stealth] (-2,12) -- (-2,13) node[left] {$m$};
\foreach \x in {2, 3, 4, 5, 6, 7, 8, 9, 10, 11}
    \draw (-2.1, \x) -- (-1.9, \x);
    
\end{tikzpicture}
    \caption{Visualisation of the \textsc{Intermediate Problem 3} for some fixed $\Phi$ and $\bfx$ and for various $k$ and $m$. As before we do not expect an answer for this problem in the unmarked regions. The restriction to the diagonal $k=m$ corresponds to the \textsc{Gapped-Relevant-Input} problem (cf.\ \cref{fig:gapped_problem}).}
    \label{fig:double_gapped_problem}
\end{figure}
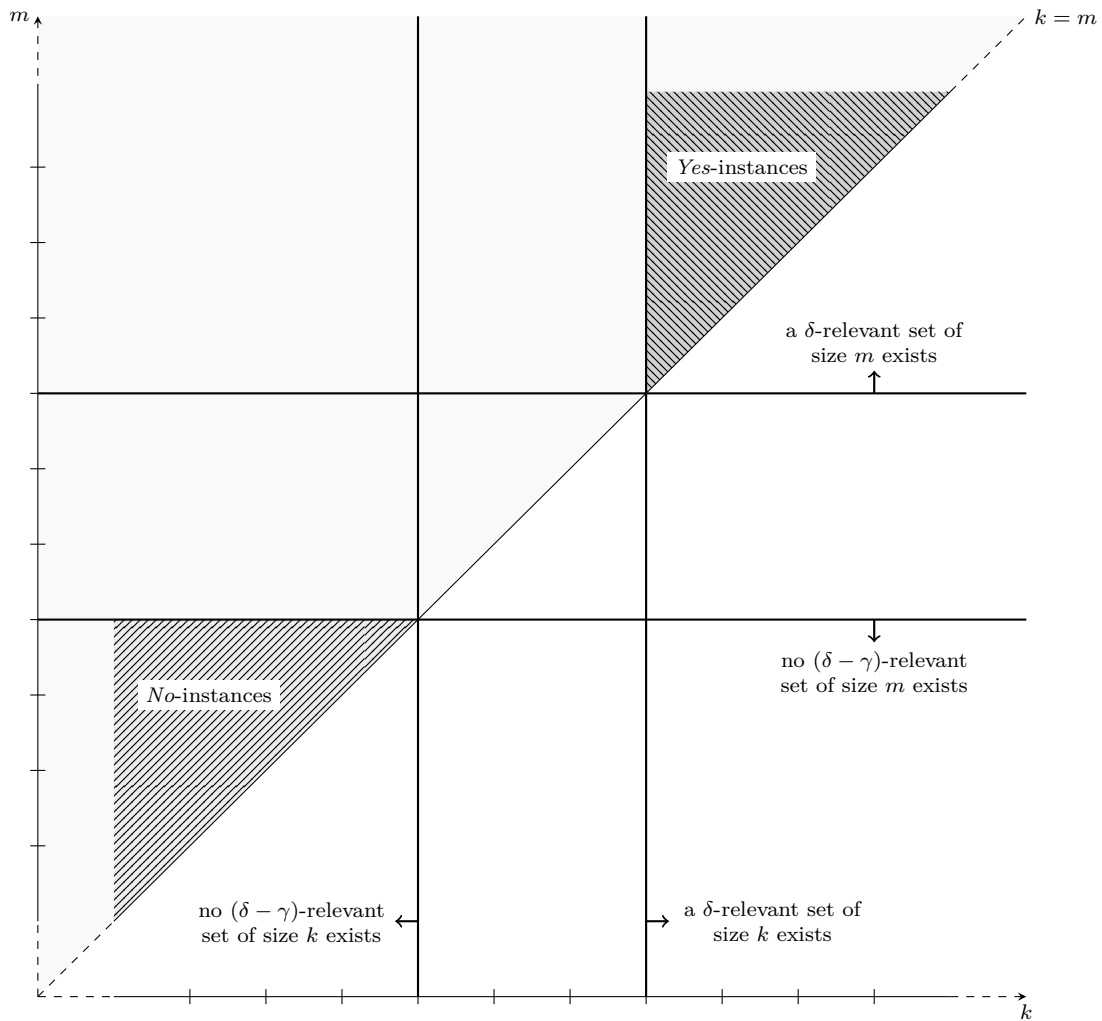

An algorithm that always produces the trivial approximate solution $k=d$ achieves an approximation factor $d$. We will show that it is generally hard to obtain better factors. More precisely, for any $\alpha>0$ an algorithm achieving an approximation factor $d^{1-\alpha}$ can not be in polynomial time unless $\SP=\SNP$.  
\begin{definition}
For $\delta \in (0,1]$ and $\gamma \in [0,\delta)$ we define the \textsc{Intermediate Problem 3} (IP3) as follows.
\begin{description}
  \item[Given:] $\Phi\colon\skl{0,1}^d\to\skl{0,1}$, $\bfx\in\skl{0,1}^d$, and $k,m\in\N$, $1\leq k\leq m\leq d$.
  \item[Decide:]~
      \begin{itemize}
	     \item[\emph{Yes}:] There exists $S\subseteq\ekl{d}$ with $|S|\leq k$ and $S$ is $\delta $-relevant for $\Phi$ and $\bfx$.
	     \item[\emph{No}:] All $S\subseteq\ekl{d}$ with $|S|\leq m$ are not $(\delta-\gamma)$-relevant for $\Phi$ and $\bfx$.
    \end{itemize}
\end{description}
\end{definition}
The restriction to the case $k=m$ is exactly the \textsc{Gapped-Relevant-Input} problem. However here we also allow the case $k<m$ with a gap in the set sizes. This is illustrated in \cref{fig:double_gapped_problem}.

\begin{lemma}\label{lem:ip3-np-hard}
  For $\delta\in(0,1)$ and $\gamma\in[0,\delta)$ we have $\textsc{SAT}\preceq_p\textsc{IP3}$, in particular, in this case \textsc{IP3} is $\SNP$-hard.
\end{lemma}
\begin{proof}
  Let $\Phi\colon\skl{0,1}^d\to\skl{0,1}$ be a \textsc{SAT} instance. We will construct $\skl{\Phi^\prime, \bfx^\prime, k^\prime, m^\prime}$ that is \emph{Yes}-instance for \textsc{IP3} if and only if $\Phi$ is a \emph{Yes}-instance for \textsc{Sat}. Let 
  \[
  q=\left\lceil\log_2\kl{\frac{d}{1-\delta}}\right\rceil\quad\text{and}\quad p=\left\lfloor\log_2\kl{\frac{1}{\delta-\gamma}}\right\rfloor+1.
  \]
  We set
  \begin{itemize}
    \item $k^\prime = dq$,
    \item $m^\prime \geq k^\prime$ arbitrary but at most polynomial in $d$,
    \item $\Phi^\prime\colon\skl{0,1}^{d\times q}\times\skl{0,1}^{m^\prime+p}\to\skl{0,1}$ with
    \[
    \Phi^\prime(\bfu^{(1)},...,\bfu^{(q)},\bfv)= \Phi\kl{\Und_{j=1}^{q}\bfu^{(j)}}\odr\kl{\Und_{i=1}^{m^\prime+p} v_i},
    \]
    where each $\bfu^{(j)}\in\skl{0,1}^d$ and the conjunction within $\Phi$ is understood component-wise,
   \item $\bfx^\prime = \bfone_{dq+m^\prime+p}$.
  \end{itemize}
  This is a polynomial time construction.  By the choice of $\Phi^\prime$ and $\bfx^\prime$ we guarantee $\Phi^\prime(\bfx^\prime)=1$ regardless of the satisfiability of $\Phi$.
  
  \paragraph{Necessity:} Let $\Phi$ be a \emph{Yes}-instance for \textsc{SAT}. This means that there exists $\bfx\in\skl{0,1}^d$ with $\Phi(\bfx)=1$. Let $S=\skl{\,i\in [d]\,:\, x_i=1\,}$ and $S^\prime = S \times [q]$. Then $\bkl{S^\prime} \leq k^\prime$. Denote
  \[
    A(\bfu^{(1)},\dots,\bfu^{(q)}) = \Und_{(i,j)\in S^\prime} u^{(j)}_i.
  \]
  Hence, $S^\prime$ is $\delta$-relevant for $\Phi^\prime$ and $\bfx^\prime$ if $P\kl{\Phi^\prime(\bfu^{(1)},\dots,\bfu^{(q)},\bfv)\,\middle|\, A(\bfu^{(1)},\dots,\bfu^{(q)})}\geq \delta$. We have
  \begin{align*}
    P\kl{\Phi^\prime(\bfu^{(1)},\dots,\bfu^{(q)},\bfv)\,\middle|\, A(\bfu^{(1)},\dots,\bfu^{(q)})} &\geq P\kl{\Phi\kl{\Und_{j=1}^{q} \bfu^{(j)}}\,\middle|\, A(\bfu^{(1)},\dots,\bfu^{(q)})}\\
    &\geq P\kl{\Und_{j=1}^{q} \bfu^{(j)}=\bfx\,\middle|\, A(\bfu^{(1)},\dots,\bfu^{(q)})}.
  \end{align*}
  From this, with a union bound, we obtain
  \begin{align*}
      P\kl{\Und_{j=1}^{q} \bfu^{(j)}=\bfx\,\middle|\, A(\bfu^{(1)},\dots,\bfu^{(q)})} &= 1-P\kl{\lnot\Und_{j=1}^{q} \bfu^{(j)}=\bfx\,\middle|\, A(\bfu^{(1)},\dots,\bfu^{(q)})} \\
      &=1-P\kl{\exists i\in S^c\,:\,\Und_{j=1}^{q} u^{(j)}_i}\\
      &\geq 1 - |S^c|2^{-q}\\
      &\geq \delta,
  \end{align*}
  which shows that $\skl{\Phi^\prime,\bfx^\prime,k^\prime,m^\prime}$ is a \emph{Yes}-instance for \textsc{IP3}.

  \paragraph{Sufficiency:} Now, conversely, let $\Phi$ be a \emph{No}-instance for \textsc{SAT}. Then for any subset $S^\prime\subseteq\ekl{dq+m^\prime+p}$ with $|S^\prime|\leq m^\prime$ we have
  \begin{align*}
  P_{\bfy^\prime}\kl{\Phi^\prime(\bfy^\prime)=\Phi^\prime(\bfx^\prime)\,\middle|\, \bfy^\prime_{S^\prime}=\bfx^\prime_{S^\prime}} &= P_{\bfy^\prime}\kl{\Phi^\prime(\bfy^\prime)\,\middle|\, \bfy^\prime_{S^\prime}=\bfone} \\
  &= P_{(\bfu^{(1)},\dots,\bfu^{(q)},\bfv)}\kl{\Und_{i=1}^{m^\prime+p} v_i\,\middle|\, (\bfu^{(1)},\dots,\bfu^{(q)},\bfv)_{S^\prime}=\bfone} \\
  &\leq 2^{-(m^\prime+p-|S^\prime|)} \\
  &\leq 2^{-p} \\
  &< \delta-\gamma.
  \end{align*}
  This shows that $S^\prime$ is not $(\delta-\gamma)$-relevant for $\Phi^\prime$ and $\bfx^\prime$, hence $\skl{\Phi^\prime,\bfx^\prime,k^\prime,m^\prime}$ is a \emph{No}-instance for \textsc{IP3}.
\end{proof}

Finally, we come to the second main theorem of the paper which shows the inapproximability of the \textsc{Min-Gapped-Relevant-Input} problem.
\begin{theorem}\label{thm:inapprox}
    Let $\delta\in(0,1)$ and $\gamma\in [0,\delta)$. Then for any $\alpha\in(0,1)$ there is no polynomial time approximation algorithm for \textsc{Min-Gapped-Relevant-Input} with an approximation factor of $d^{1-\alpha}$ unless $\SP=\SNP$.
\end{theorem}
\begin{proof}
    We prove this by showing that the existence of such an approximation algorithm would allow us to decide \textsc{IP3} in polynomial time for certain instances. These can be chosen as in the proof of \cref{lem:ip3-np-hard}, which in turn implies that we could decide \textsc{SAT} in polynomial time. This is only possible if $\SP=\SNP$. 
    
    Let $\Phi\colon\skl{0,1}^d\to\skl{0,1}$ be a \textsc{SAT} instance and $\skl{\Phi^\prime,\bfx^\prime,k^\prime,m^\prime}$ an equivalent \textsc{IP3} instance as in the proof of \cref{lem:ip3-np-hard}. We have seen that there is some freedom in the choice of $m^\prime$ as long as it satisfies $k^\prime\leq m^\prime$ and is at most polynomial in $d$.
    We choose $ m'=\left\lceil \max(2k'({k'}^{1-\alpha}+p^{1-\alpha}), (2k')^{\frac{1}{\alpha}}+1) \right\rceil $ with $p=\left\lfloor\log_2\kl{\frac{1}{\delta-\gamma}}\right\rfloor+1$ as before. Recall that $k^\prime = dq$ with $q=\left\lceil\log_2\kl{\frac{d}{1-\delta}}\right\rceil$, so clearly $m^\prime$ is polynomial in $d$ and $k^\prime\leq m^\prime$. Further, we have $ m'> (2k')^{\frac{1}{\alpha}} $ so $ 1-k'{m'}^{-\alpha}>\frac{1}{2} $ and therefore
    \[ m'(1-k'{m'}^{-\alpha})>\frac{m'}{2}\ge k'({k'}^{1-\alpha}+p^{1-\alpha}). \]
    Now let $d^\prime=k^\prime+m^\prime+p$ denote the number of variables of $\Phi^\prime$. By the subadditivity of the map $z\mapsto z^{1-\alpha}$ we finally obtain
    \[
        k^\prime {d^\prime}^{1-\alpha} = k^\prime(k^\prime+m^\prime+p)^{1-\alpha}
        \leq k^\prime\left({k^\prime}^{1-\alpha} + {m^\prime}^{1-\alpha} + p^{1-\alpha}\right)
        < m^\prime.
    \]
    It remains to show that an \textsc{IP3} instance with $m^\prime > k^\prime {d^\prime}^{1-\alpha}$ can be decided by an approximation algorithm for \textsc{Min-Gapped-Relevant-Input} with approximation factor ${d^\prime}^{1-\alpha}$. Assume such an algorithm exists and let $k$ be an approximate solution. Then there exists a true solution $\widetilde{k}$ with $\widetilde{k}\leq k\leq {d^\prime}^{1-\alpha}\widetilde{k}$. 
    
    Firstly, assume that $\skl{\Phi^\prime, \bfx^\prime, k^\prime, m^\prime}$ is a \emph{Yes}-instance for \textsc{IP3}. Then there is a $\delta$-relevant set of size $k^\prime$. But no set smaller than $\widetilde{k}$ can be $\delta$-relevant. This implies $\widetilde{k}\leq k^\prime$ and therefore $k\leq{d^\prime}^{1-\alpha}k^\prime <m^\prime$.
    
    Secondly, assume that $\skl{\Phi^\prime, \bfx^\prime, k^\prime, m^\prime}$ is a \emph{No}-instance for \textsc{IP3}. Then all sets of size at most $m^\prime$ are not $(\delta-\gamma)$-relevant. But there exists a $(\delta-\gamma)$-relevant set of size $\widetilde{k}$. This implies $k\geq \widetilde{k}>m^\prime$.
    
    Altogether, checking whether $k<m^\prime$ or $k>m^\prime$ decides $\skl{\Phi^\prime,\bfx^\prime,k^\prime, m^\prime}$.
\end{proof}

\section{Conclusion}

We showed that the task of identifying the relevant components of an input assignment to the variables of a Boolean function is complete for the complexity class $\SNP^{\SPP}$ and thus related in difficulty to the problem of planing under uncertainty. We showed furthermore that even relaxing the problem by introducing a probability gap promise that allows for efficient estimation of the fraction of assignments satisfying the Boolean function does not significantly simplify the problem. It remains hard to only approximately solve this gapped problem in any reasonable sense (beyond the trivial solution) unless $\SP = \SNP$.

We want to stress that this is a worst-case analysis and does not imply that the task is impossible  in practical applications. Many non-linear optimisation problems are $\SNP$-hard in general and yet performed successfully on a regular basis. But it strongly suggests that anything beyond heuristic solution strategies requires further restrictions on the problem.

In fact, in a companion paper \cite{mwhk-2019-rate-dist} we present a heuristic algorithm for a continuous (non-discrete) variant of the \textsc{Relevant-Input-Problem} and classifier functions with compositional (layered) structure (such as neural networks). For several image classification tasks we demonstrate numerically that our algorithm approximates small relevant sets better than widely-used comparable methods.

\section*{Acknowledgements}
The authors would like to thank Philipp Petersen for several fruitful discussions during the early stage of the project as well as Peter B\"{u}rgisser for helpful feedback on the first draft. J. M. and S. W. acknowledge support by DFG-GRK-2260 (BIOQIC). S. H. is grateful for support by CRC/TR 109 ``Discretization in Geometry and Dynamics''.  G. K. acknowledges partial support by the Bundesministerium f{\"u}r Bildung und Forschung (BMBF) through the ``Berliner Zentrum f{\"u}r Machinelles Lernen'' (BZML), by the Deutsche Forschungsgemeinschaft (DFG) through Grants CRC 1114 ``Scaling Cascades in Complex Systems'', CRC/TR 109 ``Discretization in Geometry and Dynamics'', DFG-GRK-2433 (DAEDALUS), DFG-GRK-2260 (BIOQIC), SPP 1798 ``Compressed Sensing in Information Processing'' (CoSIP), by the Berlin Mathematics Research Centre MATH+, and by the Einstein Foundation Berlin. 

\appendix

\section{Raising the Probability}\label{apx:prob_transform}
We give a constructive proof of \cref{lem:prob_transform}. 
Let $\Phi\colon\skl{0,1}^d\to\skl{0,1}$ be arbitrary and $0<\delta_1<\delta_2<1$. We will construct a monotone function $\Pi\colon\skl{0,1}^n\to\skl{0,1}$ such that $\Pi(\bfzero_n)=0$, $\Pi(\bfone_n)=1$, and
\begin{equation}\label{eq:prob_raise}
    P_\bfy\kl{\Phi(\bfy)} > \delta_1 \quad\Longleftrightarrow\quad P_{(\bfy,\bfr)}\kl{\Phi(\bfy)\odr\Pi(\bfr)} \geq \delta_2,
\end{equation}
with 
\[
n\in\mathcal{O}\kl{\kl{d+\log_2\kl{\frac{1-\delta_1}{1-\delta_2}}}^2+\log_2\kl{\frac{1}{\delta_2-\delta_1}}}.
\]
Denote $\Phi^\prime\colon\skl{0,1}^d\times\skl{0,1}^n\to\skl{0,1}\colon (\bfy,\bfr)\mapsto \Phi(\bfy)\odr\Pi(\bfr)$, then
\begin{equation} \label{eq:or_prob}
 P(\Phi^\prime) = P(\Phi) + (1-P(\Phi)) P(\Pi),
\end{equation}
which is monotonously increasing in both $P(\Phi)$ and $P(\Pi)$. Thus it suffices to consider the edge case when $P(\Phi)$ is close to $\delta_1$. Since $P(\Phi)$ can only take values in $\skl{\frac{0}{2^d}, \frac{1}{2^d},\dots,\frac{2^d}{2^d}}$ we see that \eqref{eq:prob_raise} is equivalent to the two conditions
\begin{alignat*}{2}
    P(\Phi) &= \frac{\floor{\delta_1 2^d}}{2^d} &\quad\Longrightarrow\quad P(\Phi^\prime) &< \delta_2, \\
    P(\Phi) &= \frac{\floor{\delta_1 2^d} + 1}{2^d} &\quad\Longrightarrow\quad P(\Phi^\prime) &\geq \delta_2,
\end{alignat*}
which together with \eqref{eq:or_prob} is equivalent to
\begin{align}
    \frac{\floor{\delta_1 2^d}}{2^d} + \frac{2^d - \floor{\delta_1 2^d}}{2^d} P(\Pi) &< \delta_2 \label{eq:ubound}\\
     \frac{\floor{\delta_1 2^d} + 1}{2^d} + \frac{2^d - \floor{\delta_1 2^d }-1}{2^d} P(\Pi) &\geq \delta_2\label{eq:lbound}.
\end{align}
In the case $\delta_1<\delta_2\leq\frac{\floor{\delta_1 2^d}+1}{2^d}$ the condition \eqref{eq:lbound} is always fulfilled. Further, we have $ \frac{\floor{\delta_1 2^d}}{2^d} + \frac{2^d - \floor{\delta_1 2^d}}{2^d} P(\Pi)<\delta_1+P(\Pi) $ so that if $ P(\Pi) <\delta_2-\delta_1$ also \eqref{eq:ubound} holds. Therefore, we choose
\[ \Pi(\bfr):=\bigwedge_{i=1}^n r_i\quad \text{and}\quad n:=\Bigg\lfloor\log_2\kl{\frac{1}{\delta_2-\delta_1}}\Bigg\rfloor+1. \]
If otherwise $ \delta_2>\frac{\floor{\delta_1 2^d}+1}{2^d} $ rearranging \eqref{eq:ubound} and \eqref{eq:lbound} finally gives us the bounds
\begin{equation}\label{eq:prob_pi_cond}
  a \leq P(\Pi) < b
\end{equation}
on $P(\Pi)$ where
\begin{align*}
    a &= \frac{\delta_2 2^d - \floor{\delta_1 2^d} -1}{2^d - \floor{\delta_1 2^d} -1}, \\
    b &= \frac{\delta_2 2^d - \floor{\delta_1 2^d}}{2^d - \floor{\delta_1 2^d}}.
\end{align*}
It is not hard to check that indeed we have $0\leq a < b \leq 1$. In \cref{apx:prob_func} we show for $\eta\in\ekl{0,1}$ and $\ell\in\N$ the existence of a monotone Boolean function $\Pi_{\eta,\ell}\colon \skl{0,1}^n \to \skl{0,1}$ such that $\Pi_{\eta,\ell}(\bfzero_n)=0$, $\Pi_{\eta,\ell}(\bfone_n)=1$, and
\[
 \bkl{P\kl{\Pi_{\eta,\ell}} - \eta} \leq 2^{-\ell}
\]
with $n\leq\frac{\ell(\ell+3)}{2}\in\CO(\ell^2)$. We conclude by choosing
\begin{align*}
    \eta &= \frac{b+a}{2}, \\
    \ell &= \floor[\Bigg]{-\log_2\kl{\frac{b-a}{2}}} + 1 = \floor[\Bigg]{-\log_2 \kl{\frac{ 2^d(1 - \delta_2) }{ \kl{2^d-\floor{\delta_1 2^d}}\kl{2^d-\floor{\delta_1 2^d}-1 } } }} + 1
\end{align*}
and setting $\Pi=\Pi_{\eta,\ell}$. Then $\Pi$ satisfies \eqref{eq:prob_pi_cond} and thus also \eqref{eq:prob_raise} and we get
\[
  \ell \in \CO\kl{d + \log_2\kl{\frac{1-\delta_1}{1-\delta_2}}}
\]
and therefore $n\in\CO(\ell^2)=\CO\kl{\kl{d + \log_2\kl{\frac{1-\delta_1}{1-\delta_2}}}^2}$.

\section{Lowering the Probability}\label{apx:prob_transform2}
We give a constructive proof of \cref{lem:prob_transform2}. Let $\Phi\colon\skl{0,1}^d\to\skl{0,1}$ be arbitrary and $0<\delta_1<\delta_2<1$. We will construct a monotone function $\Pi\colon\skl{0,1}^n\to\skl{0,1}$ such that
\begin{equation}\label{eq:prob_lower}
    P_\bfy\kl{\Phi(\bfy)} \geq \delta_2 \quad\Longleftrightarrow\quad P_{(\bfy,\bfr)}\kl{\Phi(\bfy)\und\Pi(\bfr)} > \delta_1,
\end{equation}
with 
\[
n\in\CO\kl{\kl{d+\log_2\kl{\frac{\delta_2}{\delta_1}}}^2}.
\]
Denote $\Phi^\prime\colon\skl{0,1}^d\times\skl{0,1}^n\to\skl{0,1}\colon (\bfy,\bfr)\mapsto \Phi(\bfy)\und\Pi(\bfr)$, then
\begin{equation} \label{eq:and_prob}
 P(\Phi^\prime) = P(\Phi) P(\Pi),
\end{equation}
which is monotonously increasing in both $P(\Phi)$ and $P(\Pi)$. Thus it suffices to consider the edge case when $P(\Phi)$ is close to $\delta_2$. Since $P(\Phi)$ can only take values in $\skl{\frac{0}{2^d}, \frac{1}{2^d},\dots,\frac{2^d}{2^d}}$ we see that \eqref{eq:prob_lower} is equivalent to the two conditions
\begin{alignat*}{2}
    P(\Phi) &= \frac{\ceil{\delta_2 2^d}}{2^d} &\quad\Longrightarrow\quad P(\Phi^\prime) &> \delta_1, \\
    P(\Phi) &= \frac{\ceil{\delta_2 2^d} - 1}{2^d} &\quad\Longrightarrow\quad P(\Phi^\prime) &\leq \delta_1,
\end{alignat*}
which together with \eqref{eq:and_prob} is equivalent to
\begin{align}
    \frac{\ceil{\delta_2 2^d}}{2^d} P(\Pi) &> \delta_1 \label{eq:ubound2} \\
     \frac{\ceil{\delta_2 2^d} - 1}{2^d} P(\Pi) &\leq \delta_1. \label{eq:lbound2}
\end{align}
In the case $\frac{\ceil{\delta_2 2^d}-1}{2^d}\leq\delta_1<\delta_2$, notice that we can simply choose $\Pi\equiv 1$. Thus, from now on we assume that this is not the case. Then, rearranging \eqref{eq:ubound2} and \eqref{eq:lbound2} finally yields the bounds
\begin{equation}\label{eq:prob_pi_cond2}
  a < P(\Pi) \leq b
\end{equation}
on $P(\Pi)$, where
\begin{align*}
    a &= \frac{\delta_1 2^d}{\ceil{\delta_2 2^d}}, \\
    b &= \frac{\delta_1 2^d}{\ceil{\delta_2 2^d}-1}.
\end{align*}
It is not hard to check that indeed we have $0 \leq a < b \leq 1$. In \cref{apx:prob_func}, we show for $\eta\in\ekl{0,1}$ and $\ell\in\N$ the existence of a monotone Boolean function $\Pi_{\eta,\ell}\colon \skl{0,1}^n \to \skl{0,1}$ such that $\Pi_{\eta,\ell}(\bfzero_n)=0$, $\Pi_{\eta,\ell}(\bfone_n)=1$, and
\[
 \bkl{P\kl{\Pi_{\eta,\ell}} - \eta} \leq 2^{-\ell}
\]
with $n\leq\frac{\ell(\ell+3)}{2}\in\CO(\ell^2)$. We conclude by choosing
\begin{align*}
    \eta &= \frac{b+a}{2}, \\
    \ell &= \floor[\Bigg]{-\log_2\kl{\frac{b-a}{2}}} + 1 = \floor[\Bigg]{-\log_2 \kl{\frac{ \delta_1 2^d}{ 2\ceil{\delta_2 2^d}\kl{\ceil{\delta_2 2^d}-1} }}} + 1
\end{align*}
and setting $\Pi=\Pi_{\eta,\ell}$. Then $\Pi$ satisfies \eqref{eq:prob_pi_cond2} and thus also \eqref{eq:prob_lower}. Hence, we finally conclude that
\[
  \ell \in \CO\kl{d + \log_2\kl{\frac{\delta_2}{\delta_1}}}
\]
and therefore $n\in\CO(\ell^2)=\CO\kl{\kl{d + \log_2\kl{\frac{\delta_2}{\delta_1}}}^2}$.

\section{Construction of the Functions \texorpdfstring{$\Pi_{\eta,\ell}$}{PI\_eta,l}}\label{apx:prob_func}

For $\eta\in\ekl{0,1}$ (the target probability) and $\ell\in\N$ (the accuracy) we construct a Boolean function $\Pi_{\eta,\ell}\colon\skl{0,1}^n\to\skl{0,1}$ in disjunctive normal form with $n\in\CO(\ell^2)$, $\Pi_{\eta,\ell}(\bfzero_n)=0$, $\Pi_{\eta,\ell}(\bfone_n)=1$, and \[|\eta-P(\Pi_{\eta,\ell})|\leq 2^{-\ell}.\]
If $\eta\leq 2^{-\ell}$ we can simply choose $\Pi_{\eta,\ell}(x_1,\dots,x_\ell) = \Und_{k=1}^\ell x_k$.
So from now on assume $2^{-\ell}<\eta\leq 1$. In this case we construct a sequence of functions $\Pi_i\colon\skl{0,1}^{n_i}\to\skl{0,1}$ such that $p_i=P(\Pi_i)$ is monotonely increasing and converges to $\eta$ from below. We proceed according to the following iterative procedure: Start with the constant function $\Pi_0\equiv 0$. Given $\Pi_i$ and $p_i$ we can stop and set $\Pi_{\eta,\ell}=\Pi_i$ if $|\eta-p_i|\leq 2^{-\ell}$. Otherwise we set $n_{i+1} = n_i+\Delta n_i$ with
\begin{equation}\label{eq:apx_deltan}
    \Delta n_i = \argmin \skl{\,n\in\N\,:\, p_i+(1-p_i) 2^{-n}\leq\eta\,},
\end{equation}
and
\[\Pi_{i+1}(x_1,\dots,x_{n_{i+1}}) = \Pi_i(x_1,\dots,x_{n_i}) \odr\kl{\Und_{k=n_i+1}^{n_{i+1}} x_k}.\]
Clearly, we obtain $p_{i+1} = p_i + (1-p_i) 2^{-\Delta n_i}$. We will see below that $\Delta n_i$ can not be too large and thus \eqref{eq:apx_deltan} can be efficiently computed by sequential search.
\begin{lemma}\label{lem:deltai_bound}
The sequence $\kl{p_i}_{i\in\N}$ is monotonely increasing and we have 
\[|\eta-p_{i+1}| \leq \frac{1}{2} |\eta-p_i|\]
for all $i\in\N$. In particular $|\eta-p_i|\leq 2^{-i}$ and $p_i\to \eta$ as $ i\to\infty $.
\end{lemma}
\begin{proof}
Since $0=p_0\leq\eta$ and by choice of $\Delta n_i$, we have $p_i\leq \eta$ for all $i\in\N$. Also from \eqref{eq:apx_deltan} we know that $p_i + (1-p_i) 2^{-(\Delta n_i-1)} > \eta$ since otherwise $\Delta n_i$ would be chosen smaller. Therefore
\begin{align*}
    \eta-p_{i+1}  &=  \eta -p_i -(1-p_i) 2^{-\Delta n_i} \\
    &= \eta - \frac{1}{2} p_i - \frac{1}{2}\kl{p_i + (1-p_i) 2^{-(\Delta n_i -1)}}\\
    &\leq \frac{1}{2}\kl{\eta - p_i}.
\end{align*}
The second part simply follows by repeatedly applying the above recursion $i$ times and from the fact that $\eta-p_0=\eta\leq1$.
\end{proof}
We conclude that the desired accuracy is reached after at most $\ell$ iterations in which case we stop and set $\Pi_{\eta,\ell}=\Pi_\ell$. It remains to determine how many variables need to be used in total. We first bound how many variables are added in each step.
\begin{lemma}\label{lem:deltan_bound}
For any $i\in\N$, we have $\Delta n_i < -\log_2\kl{\eta-p_i} + 1$.
\end{lemma}
\begin{proof}
As before we know $p_i + (1-p_i) 2^{-(\Delta n_i-1)} > \eta$ since otherwise $\Delta n_i$ would be chosen smaller. This implies
\begin{equation*}
    2^{-(\Delta n_i -1)} > \frac{\eta-p_i}{1-p_i} \geq \eta-p_i,
\end{equation*}
and therefore $\Delta n_i < -\log_2\kl{\eta-p_i} + 1$.
\end{proof}

This can finally be used to bound how many variables are used in total.
\begin{lemma}
The total number of variables for $\Pi_{\eta,\ell}=\Pi_\ell$ is 
\[n = n_\ell = \sum_{i=1}^\ell \Delta n_{i-1} \in \CO\kl{\ell^2}.\]
\end{lemma}
\begin{proof}
From \cref{lem:deltai_bound} we get $\eta-p_i\geq 2(\eta-p_{i+1})$ and thus $\eta-p_i\geq 2^{\ell-1-i}(\eta-p_{\ell-1})$. Without loss of generality we can assume $\eta-p_{\ell-1}\geq 2^{-\ell}$ since otherwise we can stop the iterative construction of $\Pi_{\eta,\ell}$ at $\ell-1$. Using \cref{lem:deltan_bound}, this immediately results in
\begin{align*}
    n &= \sum_{i=1}^\ell \Delta n_{i-1} \\
    &\leq \sum_{i=1}^\ell -\log_2\kl{\eta - p_{i-1}} + 1\\
    &\leq \sum_{i=1}^\ell -\log_2\kl{2^{\ell-i}(\eta-p_{\ell-1})} + 1\\
    &\leq \sum_{i=1}^\ell -\log_2\kl{2^{-i}} + 1\\
    &= \frac{\ell(\ell+1)}{2} + \ell \in \CO\kl{\ell^2}.\qedhere
\end{align*}
\end{proof}

\small
\bibliographystyle{abbrv}
\bibliography{references}
\end{document}